\RequirePackage{lineno}
\documentclass[twocolumn,aps,tightenlinefs,superscriptaddress,preprintnumbers]{revtex4}
\usepackage{graphicx,color,dcolumn,booktabs,bm}
\usepackage{longtable,lscape}
\usepackage{amsmath}
\usepackage{indentfirst}
\usepackage{epsfig}
\usepackage{feynmf}
\usepackage{epstopdf}
\usepackage{slashed}
\usepackage{cases}
\usepackage{color}
\usepackage{multirow}
\usepackage{ulem}
\usepackage{graphicx,color,dcolumn,booktabs,bm}
\usepackage{verbatim}
\usepackage[colorlinks,linkcolor=blue,anchorcolor=green,citecolor=blue]{hyperref}
\usepackage{mathrsfs}
\usepackage{rotating}
\usepackage{threeparttable}
\usepackage{dcolumn}
\usepackage{orcidlink}

\newcommand{\BR}{{\cal B}}

\newcommand{\pip}{\pi^+}
\newcommand{\pim}{\pi^-}

\newcommand{\EE}{e^+e^-}

\newcommand{\beq}{\begin{equation}}
\newcommand{\eeq}{\end{equation}}
\newcommand{\bitm}{\begin{itemize}}
\newcommand{\eitm}{\end{itemize}}

\begin{document}
\hyphenpenalty=10000


\title{\quad\\[0.1cm]\boldmath Observation of the decays $B^{+} \to \Sigma_{c}(2455)^{++} \overline{\Xi}_{c}^{-}$ 
and  $B^{0} \to \Sigma_{c}(2455)^{0} \overline{\Xi}_{c}^{0}$}

\author{M.~Abumusabh\,\orcidlink{0009-0004-1031-5425}} 
\author{I.~Adachi\,\orcidlink{0000-0003-2287-0173}} 
\author{L.~Aggarwal\,\orcidlink{0000-0002-0909-7537}} 
\author{H.~Ahmed\,\orcidlink{0000-0003-3976-7498}} 
\author{Y.~Ahn\,\orcidlink{0000-0001-6820-0576}} 
\author{H.~Aihara\,\orcidlink{0000-0002-1907-5964}} 
\author{N.~Akopov\,\orcidlink{0000-0002-4425-2096}} 
\author{S.~Alghamdi\,\orcidlink{0000-0001-7609-112X}} 
\author{M.~Alhakami\,\orcidlink{0000-0002-2234-8628}} 
\author{A.~Aloisio\,\orcidlink{0000-0002-3883-6693}} 
% ...
\collaboration{The Belle and Belle II Collaborations}

\begin{abstract}
We report the first observation of the two-body  baryonic decays $B^{+} \to \Sigma_{c}(2455)^{++} \overline{\Xi}_{c}^{-}$ and  
$B^{0} \to \Sigma_{c}(2455)^{0} \overline{\Xi}_{c}^{0}$ with significances of $7.3\,\sigma$ and $6.2\,\sigma$, respectively, including 
statistical and systematic uncertainties.  The branching fractions are measured to be $\mathcal{B}(B^{+} \to \Sigma_{c}(2455)^{++} \overline{\Xi}_{c}^{-}) 
= (5.74 \pm 1.11 \pm 0.42_{-1.53}^{+2.47}) \times 10^{-4}$ and $\mathcal{B}(B^{0} \to \Sigma_{c}(2455)^{0} \overline{\Xi}_{c}^{0}) 
= (4.83 \pm 1.12 \pm 0.37_{-0.60}^{+0.72}) \times 10^{-4}$. The first and second uncertainties are statistical  and systematic, respectively, 
while the third ones arise from the absolute branching fractions of $\overline{\Xi}_{c}^{-}$ or $\overline{\Xi}_{c}^{0}$ decays.  
The data samples used for this analysis have integrated luminosities of 711~$\mathrm{fb}^{-1}$ and 365~$\mathrm{fb}^{-1}$, 
and were collected at the $\Upsilon(4S)$ resonance by the Belle and Belle~II detectors operating at the KEKB and SuperKEKB asymmetric-energy 
$e^{+}e^{-}$ colliders, respectively.
\end{abstract}

\maketitle
Baryonic $B$ decays provide an important dynamical system for studying the  production mechanisms of 
baryon-antibaryon pairs in the nonperturbative regime of quantum chromodynamics (QCD).
Over the past three decades, a number of such decays have been observed~\cite{ParticleDataGroup:2024cfk} that have many interesting features, such as
threshold enhancements in the baryon-antibaryon mass spectra~\cite{Belle:2002bro, Belle:2003pwf, BaBar:2005sdl, LHCb:2017obv} 
and a  hierarchy in the branching fractions between two-body and multibody decays~\cite{Chistov:2016kae, Huang:2021qld}. 
These observations help elucidate the intricate kinematic and dynamical properties of baryonic $B$ decays~\cite{BaBar:2014omp}.

In 2003, the Belle experiment reported the first observation of a  two-body baryonic decay,  and measured the decay  
$B^{0} \to \overline{\Lambda}_{c}^{-} p$ to have a branching fraction of order $10^{-5}$~\cite{Belle:2002gir}.
In 2006, Belle observed the double-charm decays $B \to \Lambda_{c}^{+} \overline{\Xi}_{c}$~\cite{Belle:2005gtu}; 
this result was later confirmed by the BaBar experiment~\cite{BaBar:2007xtc}.  The double-charm decays have branching fractions
of order $10^{-3}$. The decays $B^{0} \to \overline{\Lambda}_{c}^{-} p$ and $B \to \Lambda_{c}^{+} \overline{\Xi}_{c}$,
which proceed via the quark-level transitions $b \to cd\overline{u}$ and $b \to cs\overline{c}$, respectively, 
involve combinations of Cabibbo-Kobayashi-Maskawa (CKM) matrix elements of comparable magnitudes~\cite{Cheng:2006bn}.
Nevertheless, their branching fractions differ by nearly two orders of magnitude, 
suggesting that certain mechanisms may enhance or suppress specific processes.
Several possible mechanisms have been proposed to account for the  large decay rates into pairs of charmed baryons,
including $\sigma/\pi$ meson exchange via soft nonperturbative interactions~\cite{Cheng:2005vd, Cheng:2009yz}, final-state interactions~\cite{Chen:2006fsa}, and hard gluon exchange~\cite{Rui:2024xgc}. 
Further measurements of $B$ decays into charmed baryon pairs 
are useful for probing the underlying dynamics and discriminating among different theoretical mechanisms.

\begin{figure}[htbp]
	\centering
	{\includegraphics[width=9cm]{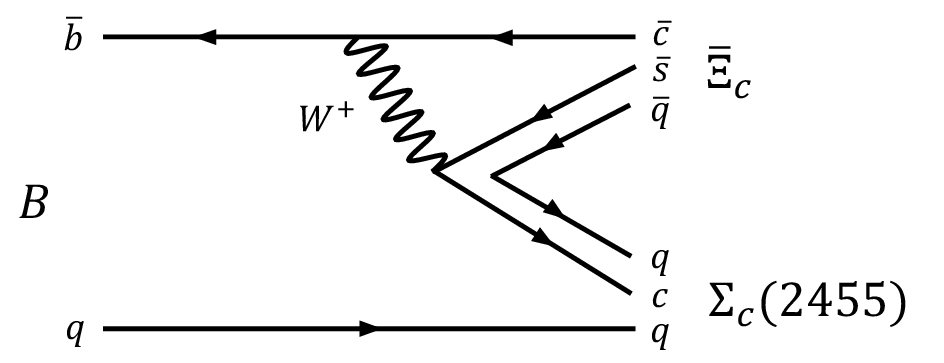}}
	\caption{Diagram representing the internal $W$-boson emission amplitude for the decays $B^{+} \to \Sigma_{c}(2455)^{++} \overline{\Xi}_{c}^{-}$ 
	and $B^{0} \to \Sigma_{c}(2455)^{0} \overline{\Xi}_{c}^{0}$, corresponding to $q = u$ and $q = d$, respectively.}\label{fig1}
\end{figure}

Theoretical studies of the  $B^{+} \to \Sigma_{c}(2455)^{++} \overline{\Xi}_{c}^{-}$ and 
$B^{0} \to \Sigma_{c}(2455)^{0} \overline{\Xi}_{c}^{0}$ decays use a QCD 
sum rule~\cite{Chernyak:1990ag} and the diquark model~\cite{Ball:1990fw}. 
The QCD sum rule predicted these double-charm branching fractions to be as large as $4 \times 10^{-3}$~\cite{Chernyak:1990ag}, while the diquark model
estimated them to be of the order $10^{-4}$, or 30\%-70\% of those of the 	
$B^{+} \to \Lambda_{c}^{+} \overline{\Xi}_{c}^{0}$ and  $B^{0} \to \Lambda_{c}^{+} \overline{\Xi}_{c}^{-}$ decays~\cite{Ball:1990fw}.
These two decays proceed through a purely internal $W$-boson emission amplitude~\cite{Hsiao:2023mud}, as shown in Fig.~\ref{fig1}.
This topology gives rise to a nonfactorizable amplitude~\cite{Cheng:2006nm}, stemming from the nonperturbative
QCD dynamics such as final-state interactions and soft gluon exchanges~\cite{Verma:1995tc, Katoch:1997mk, Kamal:1995fr, Kaur:2025pvi}.
These two decay modes thus provide a theoretically reliable environment to probe such effects. 
In addition, according to $SU(3)$  flavor symmetry, the $\Sigma_c(2455)$ baryon belongs to a sextet of flavor-symmetric states,
while the $\overline{\Xi}_c$ baryon belongs to an antitriplet of flavor-antisymmetric states.
To date, no $B$ decays into charmed baryon pairs containing both an antitriplet and a sextet have been observed.

We report the first search for the decays $B^{+} \to \Sigma_{c}(2455)^{++} \overline{\Xi}_{c}^{-}$ and
$B^{0} \to \Sigma_{c}(2455)^{0} \overline{\Xi}_{c}^{0}$. 
Charge-conjugate channels are implicitly included throughout this analysis.
This study is based on data samples that have integrated luminosities
of 711~$\rm{fb}^{-1}$~\cite{Belle:2012iwr}, collected by the Belle detector~\cite{Belle:2000cnh}, and 365~$\rm{fb}^{-1}$~\cite{Belle-II:2024vuc},
collected by the Belle~II detector~\cite{Belle-II:2010dht}, at the $e^{+}e^{-}$ center-of-mass (c.m.) energy ($\sqrt{s}$) of $10.58$~GeV.
The datasets contain $(772 \pm 11) \times 10^{6}$ $\Upsilon(4S)$ events for 
Belle and $(387 \pm 6) \times 10^{6}$ $\Upsilon(4S)$ events for Belle II.
The $\Sigma_{c}(2455)^{++, 0}$  baryons are reconstructed 
in their $\Lambda_{c}^{+} \pi^{\pm}$ decays followed by the
$\Lambda_{c}^{+} \to p K^{-} \pip$ and $\Lambda_{c}^{+} \to p K_{S}^{0}$ decays.
The $\overline{\Xi}_{c}^{-}$ baryon is reconstructed via
$\overline{\Xi}_{c}^{-} \to \overline{\Xi}^{+} \pi^{-} \pi^{-}$ and $\overline{p} K^{+} \pi^{-}$ decays, and the $\overline{\Xi}_{c}^{0}$ baryon
via $\overline{\Xi}_{c}^{0} \to \overline{\Xi}^{+} \pi^{-}$ and $\overline{\Lambda} K^{+} \pi^{-}$ decays, followed by $\overline{\Xi}^{+} \to \overline{\Lambda} \pi^{+}$.
To avoid experimental bias, the signal region is not examined until the analysis procedure is finalized.
All selection criteria are determined by iteratively optimizing the figure-of-merit for an observation at a significance level of five standard deviations based on simulation~\cite{Punzi}.
The signal yields are extracted from a two-dimensional (2D) fit to the distributions of the difference between the expected and observed $B$ 
meson energy and the $\Lambda_{c}^{+} \pi$ invariant mass. The 2D fit is performed simultaneously 
on events from the signal and sideband regions of the $\overline{\Xi}_{c}$ invariant mass.

The Belle detector operated at the KEKB~\cite{Kurokawa:2001nw} asymmetric-energy $\EE$ collider, 
while the Belle~II detector operates at its successor, the SuperKEKB collider~\cite{Akai:2018mbz}. 
The two detectors are nearly 4$\pi$ hermetic solenoidal magnetic spectrometers. 
They both consist of an inner silicon vertex detector and a central drift chamber, surrounded by Cherenkov-based 
charged-particle identification detectors, a crystal electromagnetic calorimeter, and outer detectors for muon and $K_L^0$ 
meson identification via penetration depth. Detailed descriptions of the Belle and Belle~II detectors can be found 
in Refs.~\cite{Belle:2000cnh, Belle-II:2010dht}. 

Monte Carlo (MC) simulated signal events are used to optimize the selection criteria, calculate the
reconstruction efficiencies, and determine the fit models.
The \textsc{evtgen}~\cite{Lange:2001uf} and \textsc{pythia}~\cite{Sjostrand:2006za,Sjostrand:2014zea} 
software packages are used to generate $e^+e^- \to \Upsilon(4S) \to B \overline{B}$ 
with final-state radiation simulated by the \textsc{photos} software package~\cite{Barberio:1990ms}. In the simulation, 
one $B$ meson decays inclusively,  while the other decays into a signal mode. Inclusive simulated 
samples of $e^{+} e^{-} \to q\overline{q}$, where $q$ indicates a $u$, $d$, $s$, or $c$ quark, 
and $\Upsilon(4S) \to B \overline{B}$ are used to optimize the selection criteria and identify the background sources~\cite{Zhou:2020ksj}.  
The \textsc{kkmc}~\cite{Jadach:1999vf} and \textsc{{pythia}}~\cite{Sjostrand:2006za,Sjostrand:2014zea} 
software packages are used to simulate the $e^{+} e^{-} \to q\overline{q}$ processes. 
The detector responses are modeled by the software packages \textsc{geant3}~\cite{GEANT3} for Belle
and \textsc{geant4}~\cite{GEANT4:2002zbu} for Belle II.

We use the Belle II analysis software framework (\textsc{basf2})~\cite{Kuhr:2018lps, basf2-zenodo}, a modular software toolkit 
developed for Belle~II data processing, 
to reconstruct both Belle and Belle~II data. 
The Belle data are converted into a \textsc{basf2}-compatible 
format using the \textsc{b2bii} (Belle-to-Belle~II) package~\cite{Gelb:2018agf},
enabling a unified analysis workflow across the two experiments.
The hardware trigger, which relies on total energy and neutral-particle multiplicity, 
is optimized to select hadronic events and is fully efficient for the signal modes.
In the offline analysis, the distance of closest approach to the interaction point for 
charged-particle trajectories (tracks) is required to be less than 2.0~cm in the plane perpendicular to the $z$ axis and less than 4.0~cm parallel
to it,  except for the $K_S^0$, $\overline{\Lambda}$, and $\overline{\Xi}^+$  decay products.
The $z$ axis is the solenoid axis, with positive direction along the $e^-$ beam, common to both Belle and Belle~II.
The identification of charged tracks uses the likelihood ratio
$\mathcal{R}(h|h^{\prime}) = \mathcal{L}(h)/[\mathcal{L}(h) + \mathcal{L}(h^{\prime})]$, where $\mathcal{L}(h^{(\prime)})$ 
is the likelihood of the charged track being a hadron $h^{(\prime)} =$ $p$, $K$, or $\pi$.
This likelihood ratio is determined using a particle identification (PID) algorithm that integrates information
from the Belle and Belle~II subdetectors~\cite{Nakano:2002jw, Belle-II:2025tpe}.
Tracks with $\mathcal{R}(p|K) > 0.6$ and $\mathcal{R}(p|\pi) > 0.6$ are identified as proton candidates; 
charged kaon (pion) candidates must satisfy $\mathcal{R}(K|\pi) > 0.6$ ($<0.4$). 
The efficiencies of these PID requirements range from 85\% to 94\%, with corresponding misidentification rates between 3\% and 8\%.
We omit PID requirements for the pion candidates used to reconstruct $K_{S}^{0}$, $\overline{\Lambda}$, and 
$\overline{\Xi}^+$ candidates, as their kinematic properties provide sufficient discrimination.

The $K_{S}^{0}$ candidates are first reconstructed from 
pairs of oppositely charged particles assumed to be pions with a common vertex, 
and  then selected using a neural network in Belle~\cite{Belle:2018xst} 
and a boosted decision tree in Belle~II~\cite{Belle:2021efh}. 
Both discriminators primarily rely on the kinematic information of $K_{S}^{0}$
and its decay products. The  invariant mass of $K_{S}^{0}$ candidates is required to be within $9.0$~MeV/$c^2$ of its known mass~\cite{ParticleDataGroup:2024cfk}, corresponding to approximately 2.5 times the mass resolution ($\,\sigma$).
The $\overline{\Lambda}$ candidates are reconstructed from $\overline{p} \pi^{+}$ pairs with a common vertex,
and an invariant mass within $5.5$~MeV/$c^2$ of its mass~\cite{ParticleDataGroup:2024cfk} (approximately $2.5\,\sigma$).
The selected $\overline{\Lambda}$ candidate is then combined with a $\pi^{+}$ to form a $\overline{\Xi}^+$ candidate. 
The invariant mass of $\overline{\Xi}^+$ candidates is required to be within $6.5$~MeV/$c^2$ of
its mass~\cite{ParticleDataGroup:2024cfk} (approximately $2.5\,\sigma$).


The invariant masses of the $\Lambda_c^+$, $\overline{\Xi}_c^-$, and $\overline{\Xi}_c^0$ charmed baryon candidates
are required to lie within 15.0, 18.0, and 18.0~MeV$/c^2$ of their known values~\cite{ParticleDataGroup:2024cfk},
respectively, corresponding to mass ranges of approximately 2.5$\,\sigma$.
The selected $\Lambda_c^+$ candidates are combined with $\pi^\pm$ candidates to form
$\Sigma_c(2455)^{++,0}$ candidates, which are subsequently combined with $\overline{\Xi}_c^{-,0}$ candidates to reconstruct $B^{+,0}$ candidates.
Each signal channel thus has four distinct reconstruction modes.
For each of the intermediate particle candidates ($K_{S}^{0}$, $\overline{\Lambda}$, $\overline{\Xi}^{+}$,
$\Lambda_{c}^{+}$, $\Sigma_{c}(2455)^{++, 0}$, and $\overline{\Xi}_{c}^{-, 0}$), 
the tracks associated with its decay products are fitted to a common vertex,
and the invariant mass is constrained to the corresponding known value~\cite{ParticleDataGroup:2024cfk}. 
A vertex fit is applied to the $B^{+, 0}$  candidates.
When reconstructing modes involving $\overline{\Xi}_{c}^{-} \to \overline{p} K^{+} \pi^{-}$
decays, a requirement of $\chi^2/\text{ndf}< 10$ on the $B^+$ vertex fit is imposed to 
further suppress combinatorial background, where $\rm ndf$ is the number of degrees of freedom.
If multiple candidates are present in an event, all combinations are retained for further analysis.
The fraction of events with multiple candidates ranges from 3\% to 5\% in data,
in agreement with expectations from simulation. The average number of candidates in such events
is between 2.02 and 2.06, with misreconstructed candidates contributing as smooth background.

Backgrounds are studied using both inclusive MC samples and data 
from the sideband regions of the $M(\Lambda_c^+)$, $M(\overline{\Xi}_{c}^{-,0})$, and $M_{\rm bc}$ distributions.
The $M(\Lambda_c^+)$ and $M(\overline{\Xi}_{c}^{-,0})$ denote the invariant masses of the reconstructed $\Lambda_c^+$ and $\overline{\Xi}_{c}^{-,0}$ candidates, and 
the $M_{\rm bc}$ is defined as $M_{\rm bc} = \sqrt{E_{\rm beam}^2 - \left(\sum_i \vec{p}_i\right)^2}$,
where $E_{\rm beam} = \sqrt{s}/2$ is the beam energy in the $e^{+}e^{-}$ c.m.\ system, and $\vec{p}_i$ is the momentum of the $i$th daughter of the $B$ meson.
We require $M_{\rm bc} > 5.27~{\rm GeV}/c^2$, which retains more than 97\% of the signal.
The sideband regions of $M(\Lambda_c^+)$, 
$M(\overline{\Xi}_c^{-, 0})$, and $M_{\rm bc}$  are 
$2231.0 < M(\Lambda_{c}^{+})  < 2261.0~{\rm MeV}/c^{2}$ or $2311.0 < M(\Lambda_{c}^{+})  < 2341.0~{\rm MeV}/c^{2}$,
$2398.0 < M(\overline{\Xi}_c^{-, 0})  < 2434.0~{\rm MeV}/c^{2}$ or $2504.0 <M(\overline{\Xi}_c^{-, 0}) < 2540.0~{\rm MeV}/c^{2}$,
and $5.235< M_{\rm bc} < 5.265~{\rm GeV}/c^{2}$, respectively, which are twice as wide
as the corresponding signal region.  The corresponding $M(\Lambda_c^+ \pi^\pm)$ and $\Delta E$ 
distributions from these sideband regions in the combined Belle and Belle~II data samples are presented in the supplemental material~\cite{Supp}. 
Here and throughout, $M(\Lambda_c^+ \pi^\pm)$ is the invariant mass of the $\Sigma_c(2455)^{++,0}$ candidate, and $\Delta E$ is defined as
$\Delta E = \sum_i E_i - E_{\rm beam}$, where $E_i$ is the energy of the $i$th daughter of the $B$ meson  in the $e^+e^-$ c.m.\ frame.
The $M(\Lambda_c^+)$ and $M_{\rm bc}$ sideband events have no significant peaks 
in either  the $M(\Lambda_c^+\pi^\pm)$ or $\Delta E$ distributions, while the $M(\overline{\Xi}_c^{-,0})$ 
sideband events contain small potential peaks in both distributions.

To extract the signal yields, we perform a 2D extended maximum likelihood fit to the unbinned $M(\Lambda_c^+ \pi^\pm)$ and $\Delta E$
distributions, simultaneously using four datasets: events from the signal and sideband regions of 
$M(\overline{\Xi}_c^{-,0})$ in both Belle and Belle~II data. 
The fitting functions used to model events in the signal and sideband regions of  $M(\overline{\Xi}_c^{-,0})$ are parametrized as
\begin{equation*}\label{fiting_function1}
	\begin{aligned}
		f_{1}(M, \Delta E) & = (N^{\rm sig}_{\rm ss} + 0.5 N^{\rm sbd}_{\rm ss})s_1(M) s_2(\Delta E) \\ & +
		N^{\rm bg}_{\rm sb} s_1(M) b_2(\Delta E) 
		 + N^{\rm bg}_{\rm bs} b_1(M) s_2(\Delta E)  \\ & + N^{\rm bg}_{\rm bb}  b_1(M) b_2(\Delta E)
	\end{aligned}
\end{equation*}
and
\begin{equation*}\label{fiting_function2}
	\begin{aligned}
		f_{2}(M, \Delta E) & = N^{\rm sbd}_{\rm ss} s_1(M) s_2(\Delta E) +
		N^{\rm sbd}_{\rm sb} s_1(M) b^{\prime}_2(\Delta E) \\
		& + N^{\rm sbd}_{\rm bs} b^{\prime}_1(M)
		s_2(\Delta E)  + N^{\rm sbd}_{\rm bb}  b^{\prime}_1(M) b^{\prime}_2(\Delta E), 
	\end{aligned}
\end{equation*}
respectively. Here, $s_1(M)$ and $s_2(\Delta E)$ denote the signal probability density functions (PDFs)
for the $M(\Lambda_{c}^{+} \pi^{\pm})$ and $\Delta E$ distributions, respectively, while 
$b^{(\prime)}_1(M)$ and $b^{\prime}_2(\Delta E)$ represent the corresponding background PDFs.
The factor of 0.5 in $f_{1}(M, \Delta E)$ arises from the ratio between the defined signal 
and sideband regions of  $M(\overline{\Xi}_c^{-,0})$, as the backgrounds are found to be linear.
A Breit-Wigner function convolved with a Crystal-Ball function is used for $s_1(M)$, while a double-Gaussian function with two 
different mean values is employed for $s_2 (\Delta E)$. The width of the Breit-Wigner function
is fixed to the known intrinsic width of the $\Sigma_{c}(2455)^{++,0}$~\cite{ParticleDataGroup:2024cfk},
while the other parameters of $s_1(M)$ and $s_2(\Delta E)$ are fixed to the values obtained 
from fits to the corresponding simulated signal distributions. The background components 
$b^{(\prime)}_1(M)$ and $b^{(\prime)}_2(\Delta E)$ are  modeled by first-order polynomials with free parameters.
The signal PDFs for the sideband events are the same as those used in the signal region.
The peaking backgrounds are due to inclusive $B^{+,0} \to \Sigma_c(2455)^{++,0} X$ decays,
where $X$ denotes non-$\overline{\Xi}_{c}^{-,0}$ final states, and contribute to both signal and $M(\overline{\Xi}_{c}^{-,0})$
sideband regions.
The number of signal events is denoted by $N^{\rm sig}_{\rm ss}$, the number of peaking background events 
by $N^{\rm sbd}_{\rm ss}$, and the number of combinatorial background events
in both distributions by $N^{\rm bg, sbd}_{\rm bb}$.
The yields of background contributions that peak in one distribution 
but not in the other are denoted by $N^{\rm bg, sbd}_{\rm sb}$ and $N^{\rm bg, sbd}_{\rm bs}$, corresponding to events 
that peak in the $M(\Lambda_{c}^{+} \pi^{\pm})$ and $\Delta E$ distributions, respectively. All event yields are 
free parameters in the fit, with the signal yields in the Belle and Belle~II datasets constrained according to 
the expected ratio for a common branching fraction.

\begin{figure*}[htbp]
	\begin{center}
		\hspace{-0.7cm}
		\includegraphics[width=7cm]{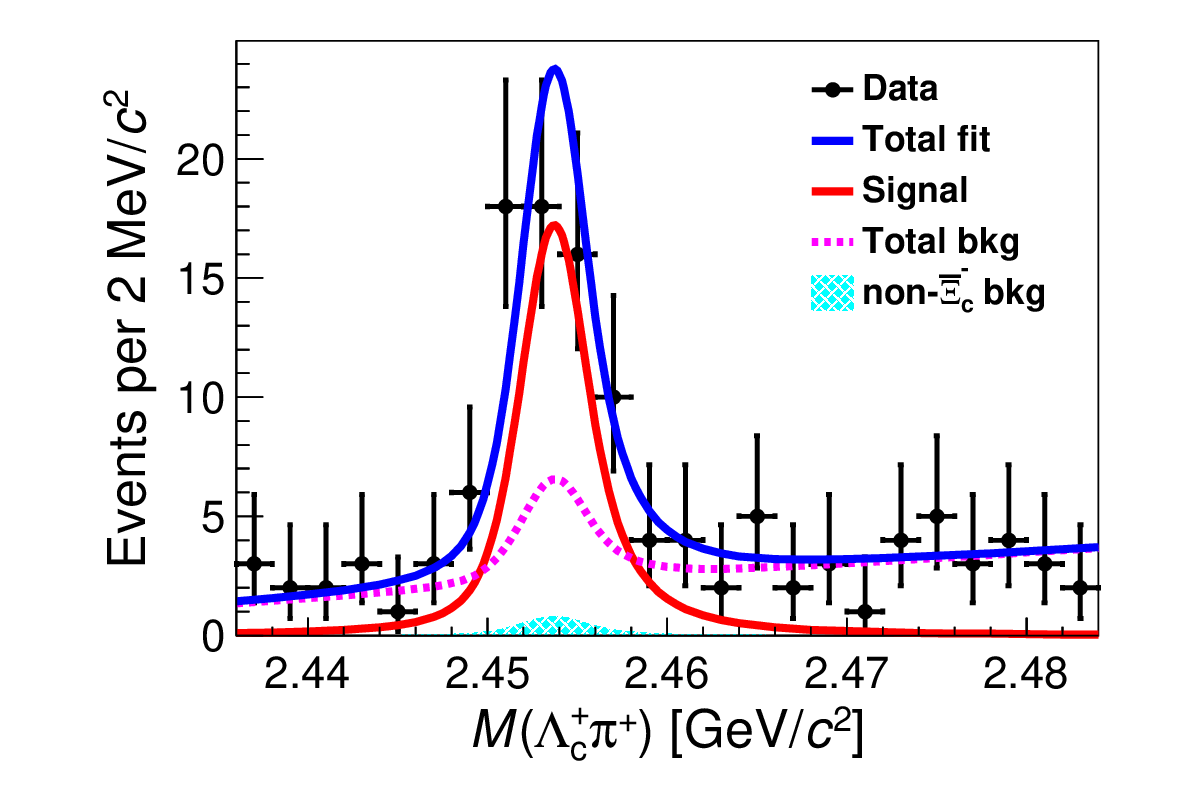}\hspace{-0.3cm}
		\includegraphics[width=7cm]{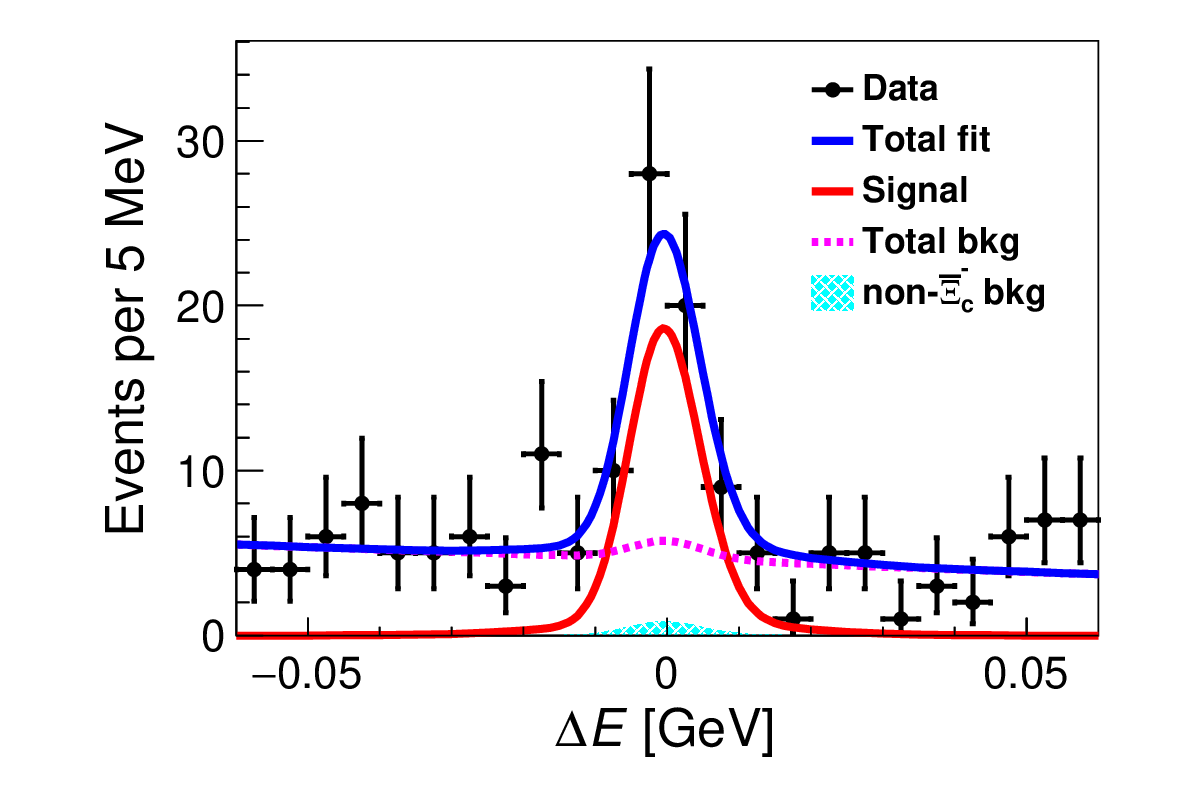}
		\put(-345,115){\bf (a)} \put(-150,115){\bf (b)}
		
		\hspace{-0.7cm}
		\includegraphics[width=7cm]{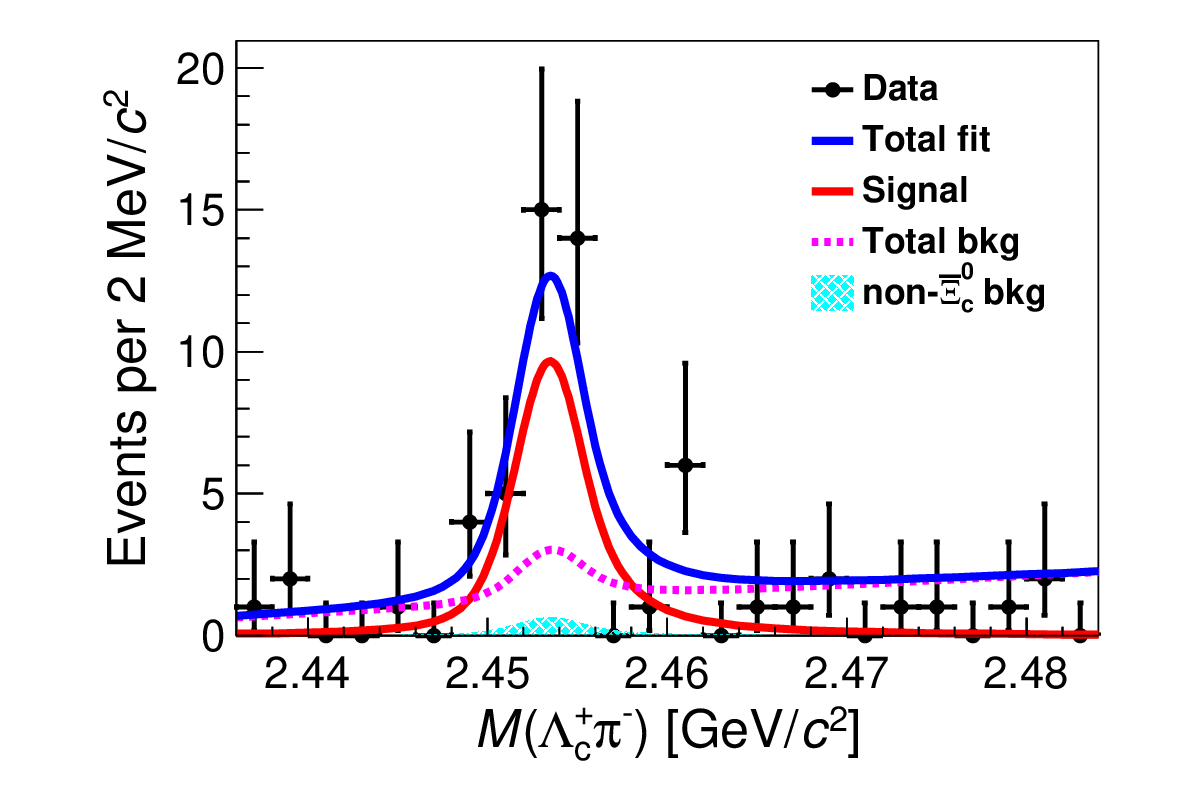}\hspace{-0.3cm}
		\includegraphics[width=7cm]{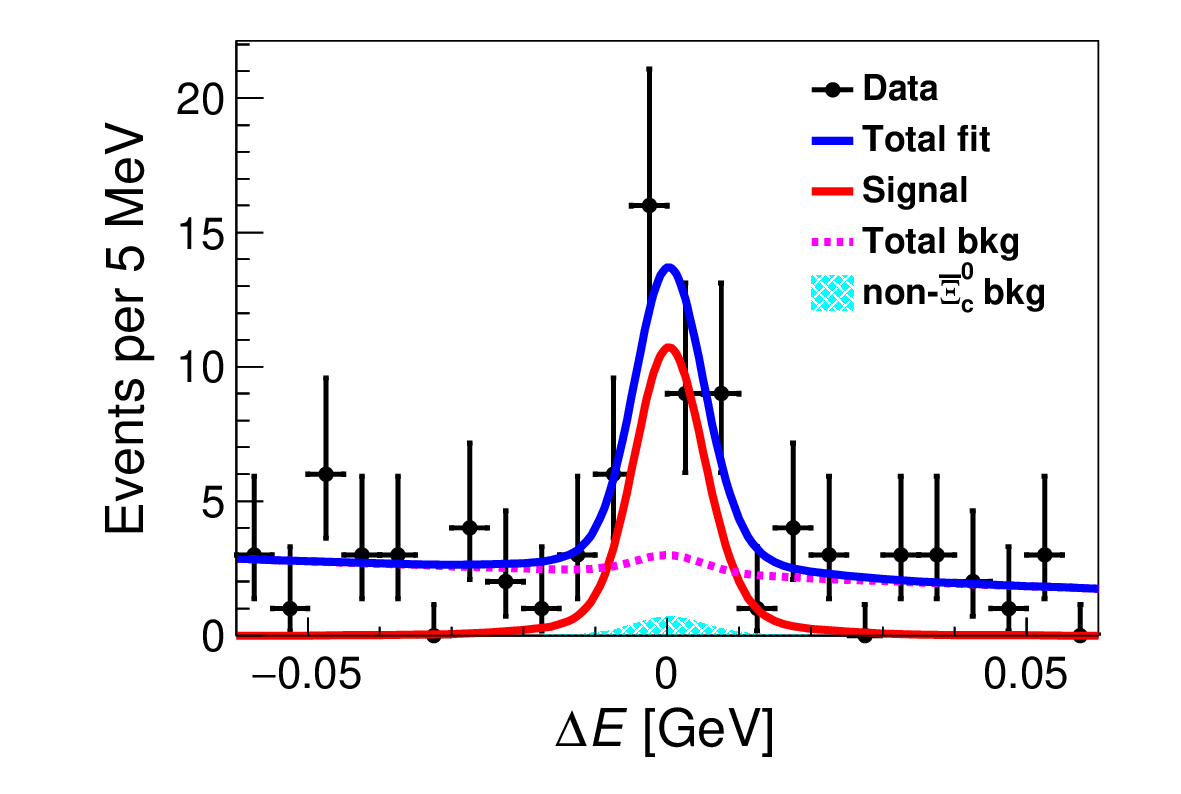}
        \put(-345,115){\bf (c)} \put(-150,115){\bf (d)}
        \caption{Distributions of (a, c) $M(\Lambda_{c}^{+} \pi^{\pm})$ and (b, d) $\Delta E$ for the reconstructed (top) $B^{+} \to \Sigma_{c}(2455)^{++} \overline{\Xi}_{c}^{-}$ and (bottom) $B^{0} \to \Sigma_{c}(2455)^{0} \overline{\Xi}_{c}^{0}$ candidates, using events from the signal regions of $M(\overline{\Xi}_{c}^{-,0})$ in the combined Belle and Belle~II datasets. Points with error bars are the data, the solid blue curves show the total fit results, the solid red curves correspond to the fitted signal components, and the dashed magenta curves represent the total fitted background components. The shaded cyan regions show the peaking-background contributions from the inclusive $B^{+,0} \to \Sigma_{c}(2455)^{++,0} X$ decays, where $X \ne \overline{\Xi}_{c}^{-,0}$.}\label{fig2}
    \end{center}
\end{figure*}

Figure~\ref{fig2} shows the $M(\Lambda_c^+ \pi^\pm)$ and $\Delta E$ distributions for events from the $M(\overline{\Xi}_c^{-,0})$ signal region in the combined Belle and Belle~II data.
Each distribution is projected within the other’s signal region, with fit results overlaid.
The signal regions for $M(\Lambda_{c}^{+} \pi^{\pm})$ and $\Delta E$  are defined as  
$2446.0 < M(\Lambda_{c}^{+} \pi^{\pm}) < 2464.0~{\rm MeV}/c^2$ and $|\Delta E| < 16~{\rm MeV}$, respectively, which retain more than 95\% of the signal.
The fitted yields of peaking backgrounds in the signal region, shown as the cyan 
components, are $2.4 \pm 3.5$ and $2.0 \pm 2.2$ for the $B^{+}$ and $B^{0}$ channels, respectively.
The corresponding fit results for events  from these sideband regions are shown in the supplemental material~\cite{Supp}.
The fitted signal yields for the decays $B^+ \to \Sigma_{c}(2455)^{++} \overline{\Xi}_{c}^{-}$ and 
$B^{0} \to \Sigma_{c}(2455)^{0} \overline{\Xi}_{c}^{0}$  are 
$52.8  \pm 10.2$ and $31.1 \pm 7.2$, respectively, with statistical significances of $7.8\,\sigma$ and $6.7\,\sigma$.
These significances are calculated using $\sqrt{-2\ln(\mathcal{L}_{0}/\mathcal{L}_\text{max})}$, 
where $\mathcal{L}_{0}$ and $\mathcal{L}_\text{max}$ are the values of the likelihoods maximized without and
with the signal component, respectively. 
To estimate the signal significances accounting for systematic uncertainties, several alternative fits are performed:  
(1) The background components $b_{1}(M)$ and $b_{2}(\Delta E)$ are modeled using either second-order polynomials or exponential functions;  
(2) The fixed signal shapes $s_1(M)$ and $s_2(\Delta E)$ are convolved with Gaussian functions that have floating resolutions;
(3) The fixed width of the $\Sigma_{c}(2455)^{++, 0}$ is varied by $\pm 1\,\sigma$~\cite{ParticleDataGroup:2024cfk};  
(4) The sideband regions of $M(\overline{\Xi}_c^{-, 0})$ are shifted by $\pm 10$~MeV/$c^2$.
Across all fit variations, the observed signal significances exceed $7.3\,\sigma$ for the $B^{+} \to \Sigma_{c}(2455)^{++} \overline{\Xi}_{c}^{-}$ 
decay and $6.2\,\sigma$ for the $B^{0} \to \Sigma_{c}(2455)^{0} \overline{\Xi}_{c}^{0}$ decay. 
These values are taken as the final signal significances after incorporating systematic effects.

The branching fractions of the $B^{+} \to \Sigma_{c}(2455)^{++} \overline{\Xi}_{c}^{-}$ and
$B^{0} \to \Sigma_{c}(2455)^{0} \overline{\Xi}_{c}^{0}$ decays are calculated using 
\begin{equation*}
	\BR = \frac{N^{\rm sig}_{\rm ss}}{ 2  f_{x} [N_{\Upsilon(4S)}^{\rm b1} \sum_{i}(\varepsilon_{i}^{\rm b1} \BR_{i}) + N_{\Upsilon(4S)}^{\rm b2} 
	\sum_{i}(\varepsilon_{i}^{\rm b2} \BR_{i})]}.
\end{equation*}
Here, $N^{\rm sig}_{\rm ss}$ represents the number of fitted $B^{+} \to \Sigma_{c}(2455)^{++} \overline{\Xi}_{c}^{-}$ or 
$B^{0} \to \Sigma_{c}(2455)^{0} \overline{\Xi}_{c}^{0}$ signal events in the combined Belle and Belle~II datasets; 
$N_{\Upsilon(4S)}^{\rm b1, b2}$ denotes the total number of $\Upsilon(4S)$ events in the Belle or Belle~II datasets;
$f_{x}$ refers to the fraction of charged ($f_{+-}$) or neutral ($f_{00}$) $B\overline{B}$ pairs~\cite{HeavyFlavorAveragingGroupHFLAV:2024ctg};
the term $\sum_{i}(\varepsilon_{i}^{\rm b1, b2} \BR_i)$ represents the sum over all reconstruction modes ($i=1-4$)
of the products of reconstruction efficiencies $\varepsilon_{i}^{\rm b1, b2}$ (for Belle or Belle~II) and the corresponding secondary branching fractions $\mathcal{B}_i$.
The numerical values of the above quantities and the calculated branching fractions are summarized in Table~\ref{table0}.
The branching fractions measured separately in Belle and Belle~II data are examined and found to be 
consistent with the results from simultaneous fits within $1\,\sigma$.

\begin{table*}[htbp]
	\centering
	\caption{\label{table0} Summary of analysis inputs and fit results. We list only the statistical uncertainties of the signal yields.
	For the branching fractions, the first and second uncertainties are statistical and systematic, respectively, while the third originates 
	from the absolute branching fractions of $\overline{\Xi}_{c}^{-, 0}$ decays~\cite{ParticleDataGroup:2024cfk}.}
	\renewcommand{\arraystretch}{1.3}
	\begin{tabular}{lccccccc}
		\hline\hline
		   & $N^{\rm sig}_{\rm ss}$ & $N^{\rm b1}_{\Upsilon(4S)}$~($10^{6}$) & $N^{\rm b2}_{\Upsilon(4S)}$~($10^{6}$) & $f_{x}$ &  $\sum_{i}(\varepsilon_i^{\rm b1} \BR_i)~(10^{-5})$
		   & $\sum_{i}(\varepsilon_i^{\rm b2} \BR_i)~(10^{-5})$ & $\BR~(10^{-4})$  \\
		    \hline  
		   $B^{+}$ &  $52.8\pm10.2$  & 772 & 387 & 0.5113  & 7.1  & 9.1  & $5.74 \pm 1.11 \pm 0.42_{-1.53}^{+2.47}$ \\
		   $B^{0}$ &  $31.1\pm7.2$    & 772 & 387 & 0.4861 & 5.2  & 6.8 & $4.83 \pm 1.12 \pm 0.37_{-0.60}^{+0.72}$ \\            
		\hline\hline 
	\end{tabular}
\end{table*}

We consider several source of systematic uncertainties,
including detection-efficiency-related (DER) uncertainties ($\sigma_{\rm DER}$), 
the statistical uncertainty on the efficiency determined from simulation ($\sigma_{\rm eff}$), 
the uncertainties on the branching fractions of 
intermediate states ($\sigma_{\BR_{i}}$), the uncertainty on the total number of $\Upsilon(4S)$ events ($\sigma_{N_{\Upsilon(4S)}}$),
the uncertainty on the fraction of charged or neutral $B\overline{B}$ events ($\sigma_{f_{x}}$), 
the possible correlation between the  $M(\Lambda_{c}^{+} \pi^{\pm})$ versus $\Delta E$ distributions
($\sigma_{\rm corr}$),  and uncertainties associated with the fit models ($\sigma_{\rm fit}$). 
Table~\ref{table1} summarizes these systematic uncertainties, with the total uncertainty ($\sigma_{\rm total}$) calculated 
as the quadratic sum of the uncertainties from each source.

The DER uncertainties include those from tracking efficiency,
PID efficiency, and the reconstructions of $K_{S}^{0}$ and $\Lambda$ candidates, which
are estimated using data control samples. The uncertainty
associated with tracking efficiency depends on the particle charge, momentum and polar angle, and ranges from
0.31\% to 0.91\% (0.38\% to 1.07\%) for each track at Belle (Belle~II), as determined from the data control samples described in Ref.~\cite{Belle:2024xcs}.
The slightly larger tracking uncertainties in Belle II result from the limited statistics of the control samples used.
The PID efficiency uncertainties are estimated to be 1.0\% (0.8\%)
for pion, 1.3\% (1.0\%) for kaon, and 2.4\% (1.8\%) for proton at Belle (Belle~II)~\cite{Nakano:2002jw, Belle-II:2025tpe}.
The $K_{S}^{0}$ reconstruction uncertainty is evaluated to be 1.2\% (1.9\%) at Belle
(Belle~II), and the $\Lambda$ reconstruction uncertainty is estimated to be 2.3\% (2.1\%). 
Both are obtained following the procedure of Ref.~\cite{Belle:2024xcs}.
The  individual uncertainties of the different modes at Belle and Belle~II are summed and
weighted by $N_{\Upsilon(4S)}^{\rm b1,b2} (\varepsilon_{i}^{\rm b1,b2} \mathcal{B}_{i})$.
Assuming these uncertainties are independent and adding them in quadrature, the detection-efficiency-related uncertainties 
are evaluated to be 2.6\%  for $B^{+} \to \Sigma_{c}(2455)^{++} \overline{\Xi}_{c}^{-}$ decay 
and 2.2\%  for $B^{0} \to \Sigma_{c}(2455)^{0} \overline{\Xi}_{c}^{0}$ decay. 
A study of the control samples $B^+ \to \Lambda_{c}^{+} \overline{\Xi}_{c}^{0}$ and 
$B^0 \to \Lambda_{c}^{+} \overline{\Xi}_{c}^{-}$, which have topologies similar to
the signal channels, indicates that the differences in vertex fit efficiencies for intermediate particles between data and simulation 
are negligible.

The statistical uncertainty of simulation-based efficiency is at most 1.0\%.
The relative uncertainties of the absolute branching fractions of $\Lambda_{c}^{+} \to p K^{-} \pip$,
$\Lambda_{c}^{+} \to p K_{S}^{0}$, $K_{S}^{0} \to \pip \pim$, $\overline{\Xi}_{c}^{-} \to
\overline{\Xi}^{+} \pim \pim$, $\overline{\Xi}_{c}^{-} \to \overline{p} K^{+} \pim$, $\overline{\Xi}_{c}^{0} \to
\overline{\Xi}^{+} \pim$, $\overline{\Xi}_{c}^{0} \to \overline{\Lambda} K^{+} \pim$, 
$\overline{\Xi}^{+} \to \overline{\Lambda} \pi^+$, and $\overline{\Lambda} \to \overline{p} \pi^+$ are
taken from Ref.~\cite{ParticleDataGroup:2024cfk}. Since the large uncertainties in the branching fractions 
of the intermediate decays $\overline{\Xi}_{c}^{-} \to \overline{\Xi}^{+} \pim \pim$ (44.8\%), $\overline{\Xi}_{c}^{-} \to \overline{p} K^{+} \pim$ (48.4\%),
$\overline{\Xi}_{c}^{0} \to \overline{\Xi}^{+} \pim$ (18.9\%), and $\overline{\Xi}_{c}^{0} \to \overline{\Lambda} K^{+} \pim$ (19.3\%)
might be reduced with future measurements, we treat them separately as a third source of uncertainty.
The branching fraction of each intermediate state is varied independently by $\pm 1\,\sigma$, 
with the resulting deviation from the nominal value taken as the corresponding systematic uncertainty.
There are uncertainties of $\ensuremath{{}_{-27\%}^{+43\%}}$, $\ensuremath{{}_{-12\%}^{+15\%}}$, and 4.0\% 
associated with the absolute branching fractions of  $\overline{\Xi}_{c}^{-}$, $\overline{\Xi}_{c}^{0}$,
and other intermediate states, respectively. 
The uncertainties of $N_{\Upsilon(4S)}^{\rm b1,b2}$ are 1.4\% for Belle~\cite{Belle:2022pcr} and 1.5\% for Belle~II~\cite{Belle:2024bsw}, 
and are combined into a total uncertainty weighted by $N_{\Upsilon(4S)}^{\rm b1,b2} \sum_{i}(\varepsilon_{i}^{\rm b1,b2} \mathcal{B}_{i})$.
The uncertainties of $f_{+-}$ and $f_{00}$ are 2.1\% and 1.7\%~\cite{HeavyFlavorAveragingGroupHFLAV:2024ctg}, respectively.
The  uncertainty arising from the possible correlation between the $M(\Lambda_c^{+}\pi^{\pm})$ and $\Delta E$  distributions is
estimated using a bootstrap method~\cite{Efron:1979bxm}.  A total of 500 bootstrap samples are constructed from the simulated samples.
For each bootstrap sample, the signal and background yields are generated by sampling from Poisson distributions centered 
at the values obtained from the fit to data. The deviation between the mean of the output signal yield distribution and the central value used
in the generation is taken as the systematic uncertainty.


The systematic uncertainties associated with the fit models arise from the empirical choice of background PDFs,
the mass resolution differences between data and simulation, the fixed width of 
$\Sigma_{c}(2455)^{++, 0}$, and the choice of sideband regions for $M(\overline{\Xi}_c^{-, 0})$. 
To estimate the uncertainty due to the background parametrization, the nominal background PDFs $b_{1}(M)$ 
and $b_{2}(\Delta E)$ are replaced with either second-order polynomials or exponential functions, and the largest 
deviation from the nominal fit result is assigned as the systematic uncertainty. 
The uncertainty due to mass-resolution differences between data and simulation is assessed by convolving 
the fixed signal shapes $s_1(M)$ and $s_2(\Delta E)$ with Gaussian functions having free widths, and the resulting deviation 
from the nominal fit is taken as the corresponding uncertainty. 
The effect of the fixed width of $\Sigma_{c}(2455)^{++, 0}$ is evaluated by varying each width by $\pm 1\,\sigma$~\cite{ParticleDataGroup:2024cfk}, 
and assigning the largest deviation as the systematic uncertainty. 
The sideband regions of $M(\overline{\Xi}_c^{-, 0})$ are shifted by $\pm 10$~MeV/$c^2$, 
and the largest resulting deviation is taken as the corresponding systematic uncertainty.
All of these contributions are summed in quadrature to obtain the total systematic uncertainty related to the fit models.

\begin{table}[htbp]
	\centering
	\caption{\label{table1} Summary of fractional systematic uncertainties (\%).}
	\vspace{0.2cm}
	\renewcommand{\arraystretch}{1.2}
	\begin{tabular}{lcccccccc}
		\hline\hline
	     & $\sigma_{\rm DER}$ & $\sigma_{\rm eff}$ & $\sigma_{\BR_{i}}$ & $\sigma_{N_{\Upsilon(4S)}}$ & $\sigma_{f_{x}}$ & $\sigma_{\rm corr}$ & $\sigma_{\rm fit}$ & $\sigma_{\rm total}$ \\
		\hline
		$B^{+}$  & 2.6 & 1.0 & 4.0 & 1.1 & 2.1 & 2.2 & 4.4 & 7.3 \\
		$B^{0}$  & 2.2 & 1.0 & 4.0 & 1.1 & 1.7 & 2.7 & 5.2  & 7.8 \\
		\hline
		\hline
	\end{tabular}
\end{table}

In summary, we report the first observation of the two-body baryonic decays
$B^{+} \to \Sigma_{c}(2455)^{++} \overline{\Xi}_{c}^{-}$ and $B^{0} \to \Sigma_{c}(2455)^{0} \overline{\Xi}_{c}^{0}$, 
using electron-positron data samples that contain
$772 \times 10^{6}$ and $387 \times 10^{6}$ $\Upsilon(4S)$ events collected by the Belle and 
Belle~II detectors, respectively. The branching fractions are measured to be 
$ \mathcal{B}(B^{+} \to \Sigma_{c}(2455)^{++} \overline{\Xi}_{c}^{-})  	= (5.74 \pm 1.11 \pm 0.42^{+2.47}_{-1.53}) \times 10^{-4} $
and $\mathcal{B}  (B^{0} \to \Sigma_{c}(2455)^{0} \overline{\Xi}_{c}^{0})  = (4.83  \pm 1.12 \pm 0.37_{-0.60}^{+0.72}) \times 10^{-4}$,
where the uncertainties are statistical, systematic, and from the absolute branching 
fractions of  $\overline{\Xi}_{c}^{-}$ or $\overline{\Xi}_{c}^{0}$ decays, respectively. 
The observed branching fractions are an order of magnitude smaller than those predicted by the QCD sum rule~\cite{Chernyak:1990ag}, 
but are consistent with the expectations of the diquark model~\cite{Ball:1990fw}.
Interestingly, these branching fractions are larger than those of their singly-charmed counterparts,
$B^{+} \to \overline{\Sigma}_{c}(2455)^{0} p$ 
and $B^{0} \to \overline{\Sigma}_{c}(2455)^{-} p$,
by one to two orders of magnitude~\cite{ParticleDataGroup:2024cfk}, although the 
corresponding combinations of CKM matrix elements in their amplitudes have nearly equal magnitudes.

We are grateful to Professors Yu-Kuo Hsiao, Hai-Yang Cheng, and Ying Li for their insightful comments and valuable suggestions.
This work, based on data collected using the Belle II detector, which was built and commissioned prior to March 2019,
and data collected using the Belle detector, which was operated until June 2010,
was supported by
Higher Education and Science Committee of the Republic of Armenia Grant No.~23LCG-1C011;
Australian Research Council and Research Grants
No.~DP200101792, 
No.~DP210101900, 
No.~DP210102831, 
No.~DE220100462, 
No.~LE210100098, 
and
No.~LE230100085; 
Austrian Federal Ministry of Education, Science and Research,
Austrian Science Fund (FWF) Grants
DOI:~10.55776/P34529,
DOI:~10.55776/J4731,
DOI:~10.55776/J4625,
DOI:~10.55776/M3153,
and
DOI:~10.55776/PAT1836324,
and
Horizon 2020 ERC Starting Grant No.~947006 ``InterLeptons'';
Natural Sciences and Engineering Research Council of Canada, Compute Canada and CANARIE;
National Key R\&D Program of China under Contract No.~2024YFA1610503,
and
No.~2024YFA1610504
National Natural Science Foundation of China and Research Grants
No.~11575017,
No.~11761141009,
No.~11705209,
No.~11975076,
No.~12135005,
No.~12150004,
No.~12161141008,
No.~12475093,
and
No.~12175041,
China Postdoctoral Science Foundation (CPSF)
under Grant No.~2024M760485, and China
Postdoctoral Fellowship Program of CPSF under Grant
No.~GZC20240303,
and Shandong Provincial Natural Science Foundation Project~ZR2022JQ02;
the Czech Science Foundation Grant No. 22-18469S,  Regional funds of EU/MEYS: OPJAK
FORTE CZ.02.01.01/00/22\_008/0004632
and
Charles University Grant Agency project No. 246122;
European Research Council, Seventh Framework PIEF-GA-2013-622527,
Horizon 2020 ERC-Advanced Grants No.~267104 and No.~884719,
Horizon 2020 ERC-Consolidator Grant No.~819127,
Horizon 2020 Marie Sklodowska-Curie Grant Agreement No.~700525 ``NIOBE''
and
No.~101026516,
and
Horizon 2020 Marie Sklodowska-Curie RISE project JENNIFER2 Grant Agreement No.~822070 (European grants);
L'Institut National de Physique Nucl\'{e}aire et de Physique des Particules (IN2P3) du CNRS
and
L'Agence Nationale de la Recherche (ANR) under Grant No.~ANR-21-CE31-0009 (France);
BMFTR, DFG, HGF, MPG, and AvH Foundation (Germany);
Department of Atomic Energy under Project Identification No.~RTI 4002,
Department of Science and Technology,
and
UPES SEED funding programs
No.~UPES/R\&D-SEED-INFRA/17052023/01 and
No.~UPES/R\&D-SOE/20062022/06 (India);
Israel Science Foundation Grant No.~2476/17,
U.S.-Israel Binational Science Foundation Grant No.~2016113, and
Israel Ministry of Science Grant No.~3-16543;
Istituto Nazionale di Fisica Nucleare and the Research Grants BELLE2,
and
the ICSC – Centro Nazionale di Ricerca in High Performance Computing, Big Data and Quantum Computing, funded by European Union – NextGenerationEU;
Japan Society for the Promotion of Science, Grant-in-Aid for Scientific Research Grants
No.~16H03968,
No.~16H03993,
No.~16H06492,
No.~16K05323,
No.~17H01133,
No.~17H05405,
No.~18K03621,
No.~18H03710,
No.~18H05226,
No.~19H00682, 
No.~20H05850,
No.~20H05858,
No.~22H00144,
No.~22K14056,
No.~22K21347,
No.~23H05433,
No.~26220706,
and
No.~26400255,
and
the Ministry of Education, Culture, Sports, Science, and Technology (MEXT) of Japan;
National Research Foundation (NRF) of Korea Grants
No.~2021R1-F1A-1064008,
No.~2022R1-A2C-1003993,
No.~2022R1-A2C-1092335,
No.~RS-2016-NR017151,
No.~RS-2018-NR031074,
No.~RS-2021-NR060129,
No.~RS-2023-00208693,
No.~RS-2024-00354342
and
No.~RS-2025-02219521,
Radiation Science Research Institute,
Foreign Large-Size Research Facility Application Supporting project,
the Global Science Experimental Data Hub Center, the Korea Institute of Science and
Technology Information (K25L2M2C3)
and
KREONET/GLORIAD;
Universiti Malaya RU grant, Akademi Sains Malaysia, and Ministry of Education Malaysia;
Frontiers of Science Program Contracts
No.~FOINS-296,
No.~CB-221329,
No.~CB-236394,
No.~CB-254409,
and
No.~CB-180023, and SEP-CINVESTAV Research Grant No.~237 (Mexico);
the Polish Ministry of Science and Higher Education and the National Science Center;
the Ministry of Science and Higher Education of the Russian Federation
and
the HSE University Basic Research Program, Moscow;
University of Tabuk Research Grants
No.~S-0256-1438 and No.~S-0280-1439 (Saudi Arabia), and
Researchers Supporting Project number (RSPD2025R873), King Saud University, Riyadh,
Saudi Arabia;
Slovenian Research Agency and Research Grants
No.~J1-50010
and
No.~P1-0135;
Ikerbasque, Basque Foundation for Science,
State Agency for Research of the Spanish Ministry of Science and Innovation through Grant No. PID2022-136510NB-C33, Spain,
Agencia Estatal de Investigacion, Spain
Grant No.~RYC2020-029875-I
and
Generalitat Valenciana, Spain
Grant No.~CIDEGENT/2018/020;
the Swiss National Science Foundation;
The Knut and Alice Wallenberg Foundation (Sweden), Contracts No.~2021.0174 and No.~2021.0299;
National Science and Technology Council,
and
Ministry of Education (Taiwan);
Thailand Center of Excellence in Physics;
TUBITAK ULAKBIM (Turkey);
National Research Foundation of Ukraine, Project No.~2020.02/0257,
and
Ministry of Education and Science of Ukraine;
the U.S. National Science Foundation and Research Grants
No.~PHY-1913789 
and
No.~PHY-2111604, 
and the U.S. Department of Energy and Research Awards
No.~DE-AC06-76RLO1830, 
No.~DE-SC0007983, 
No.~DE-SC0009824, 
No.~DE-SC0009973, 
No.~DE-SC0010007, 
No.~DE-SC0010073, 
No.~DE-SC0010118, 
No.~DE-SC0010504, 
No.~DE-SC0011784, 
No.~DE-SC0012704, 
No.~DE-SC0019230, 
No.~DE-SC0021274, 
No.~DE-SC0021616, 
No.~DE-SC0022350, 
No.~DE-SC0023470; 
and
the Vietnam Academy of Science and Technology (VAST) under Grants
No.~NVCC.05.12/22-23
and
No.~DL0000.02/24-25.

These acknowledgements are not to be interpreted as an endorsement of any statement made
by any of our institutes, funding agencies, governments, or their representatives.

We thank the SuperKEKB team for delivering high-luminosity collisions;
the KEK cryogenics group for the efficient operation of the detector solenoid magnet and IBBelle on site;
the KEK Computer Research Center for on-site computing support; the NII for SINET6 network support;
and the raw-data centers hosted by BNL, DESY, GridKa, IN2P3, INFN,
PNNL/EMSL, and the University of Victoria.

\end{document}


\hyphenpenalty=10000

\title{\quad\\[0.1cm]Supplemental  Material for ``Observation of the decays $B^{+} \to \Sigma_{c}(2455)^{++} \overline{\Xi}_{c}^{-}$ and  $B^{0} \to \Sigma_{c}(2455)^{0} \overline{\Xi}_{c}^{0}$"}

\maketitle

\begin{figure}[htbp]
	\begin{center}
		\includegraphics[width=7cm]{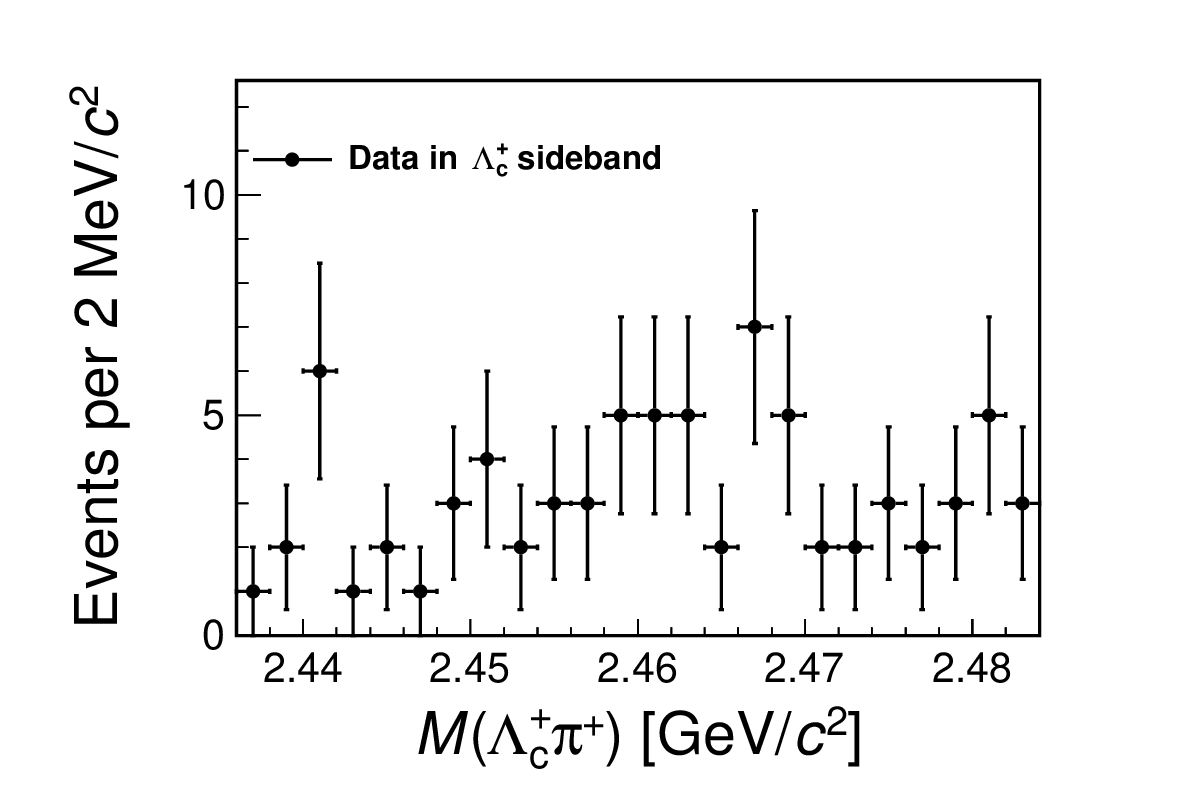}
		\includegraphics[width=7cm]{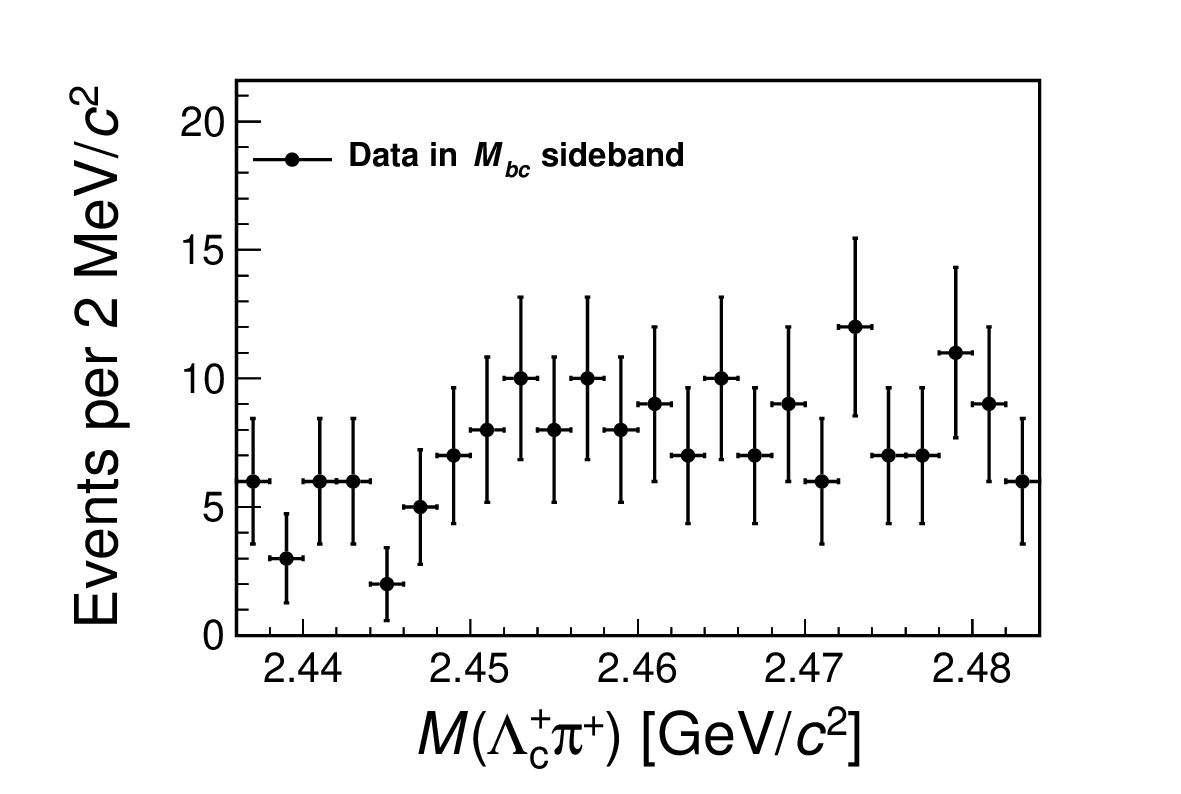}
		\put(-255,105){\bf (a1)} \put(-53,105){\bf (a2)}
	
		\includegraphics[width=7cm]{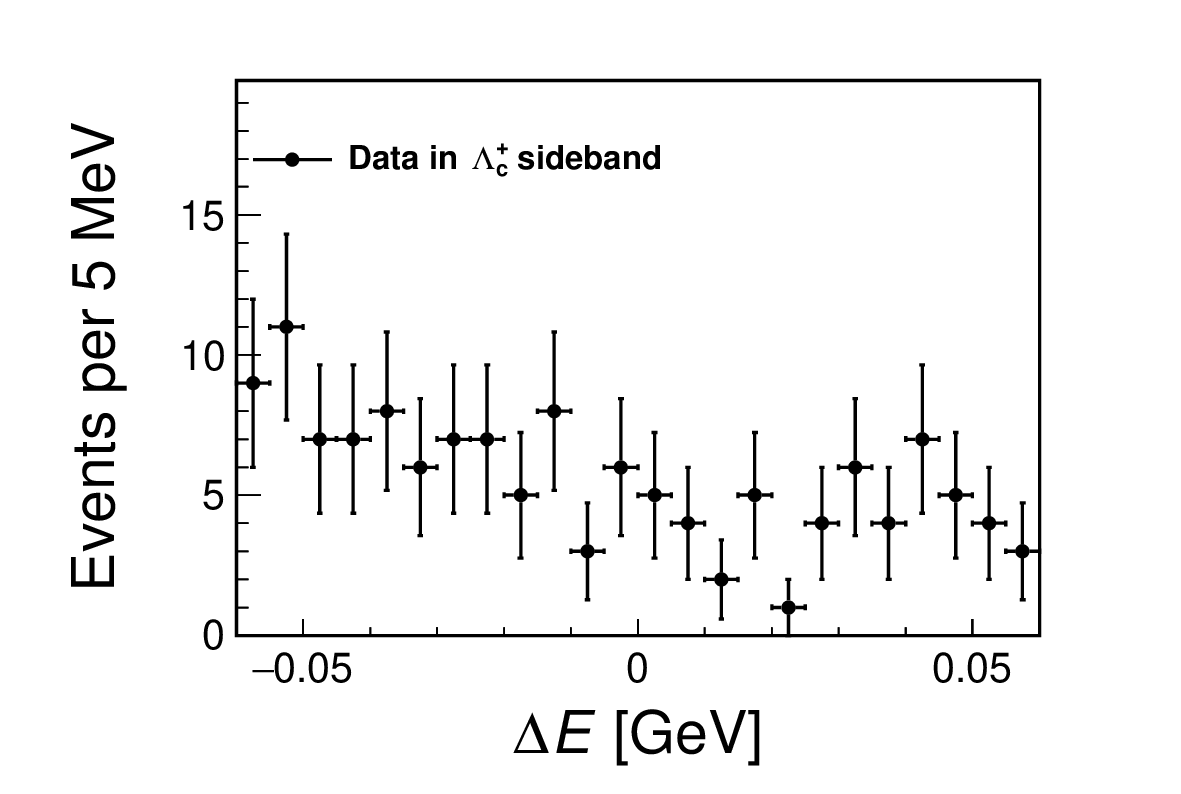}
        \includegraphics[width=7cm]{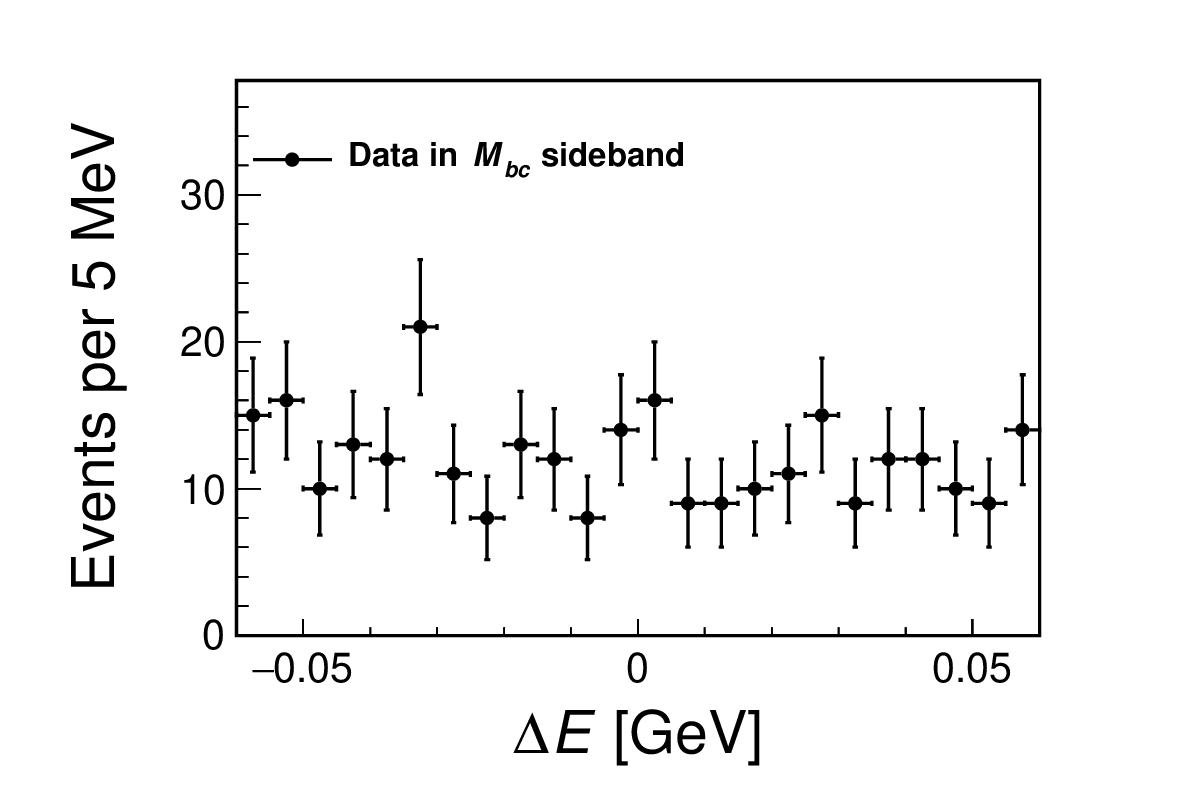}
   		\put(-255,105){\bf (b1)} \put(-53,105){\bf (b2)}
		\caption{Distributions of (a) $M(\Lambda_{c}^{+} \pi^{+})$ and (b) $\Delta E$ from the sideband regions of (1) $M(\Lambda_{c}^{+})$ and  (2) $M_{bc}$
		for the $B^{+} \to \Sigma_{c}(2455)^{++} \overline{\Xi}_{c}^{-}$ decay in the combined Belle and  Belle~II data sets.}\label{fig1}
	\end{center}
\end{figure}

\begin{figure}[htbp]
	\begin{center}
		\includegraphics[width=7cm]{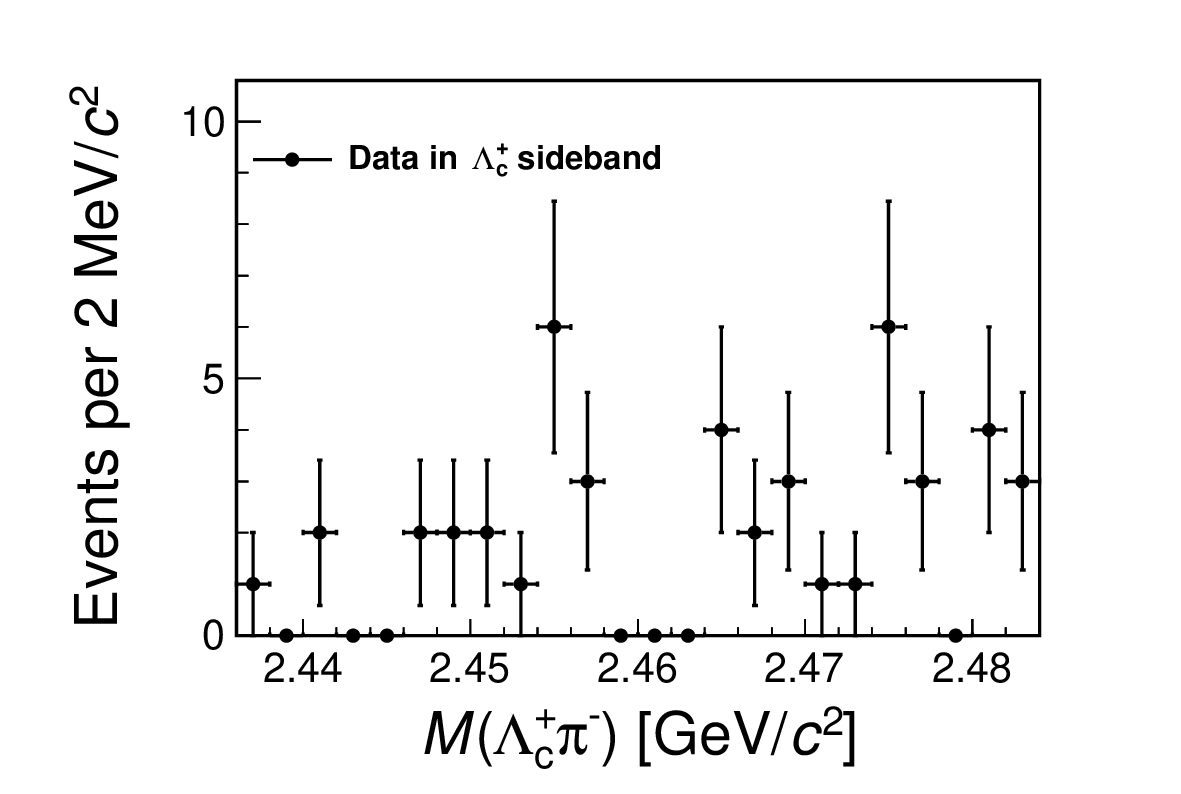}
		\includegraphics[width=7cm]{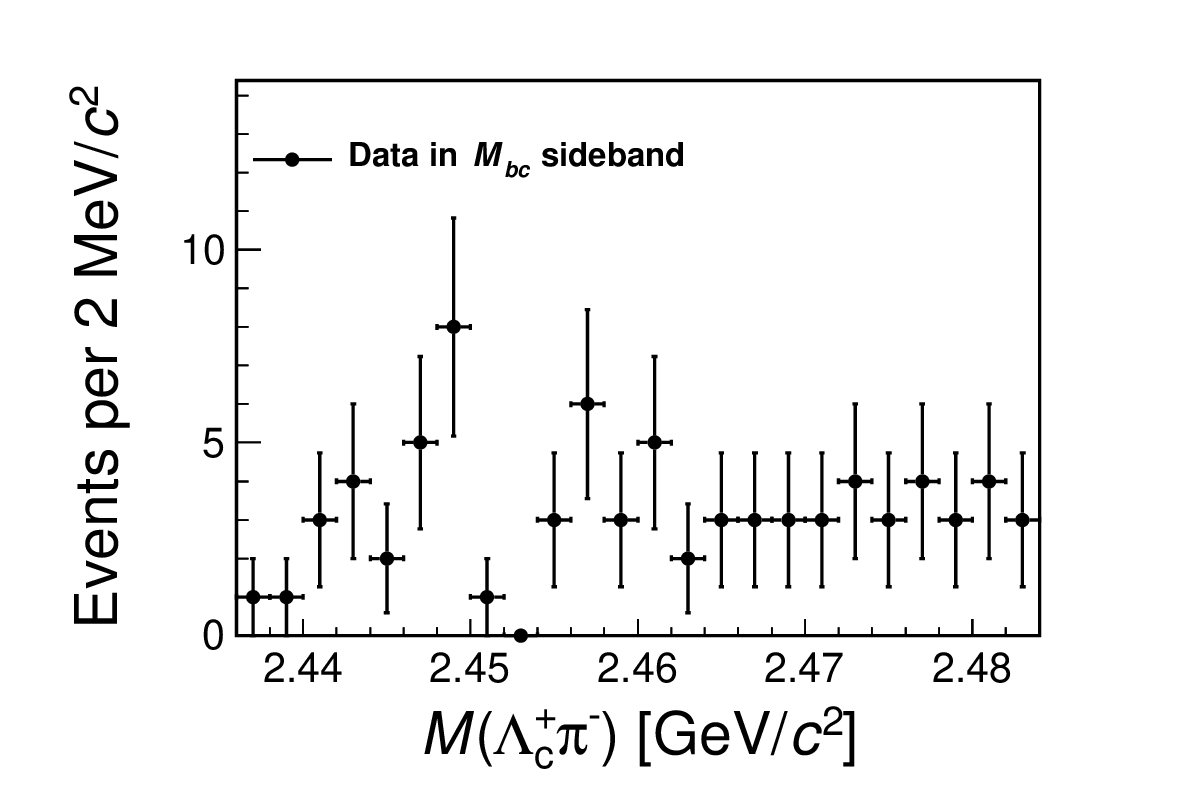}
		\put(-255,105){\bf (a1)} \put(-53,105){\bf (a2)}

		\includegraphics[width=7cm]{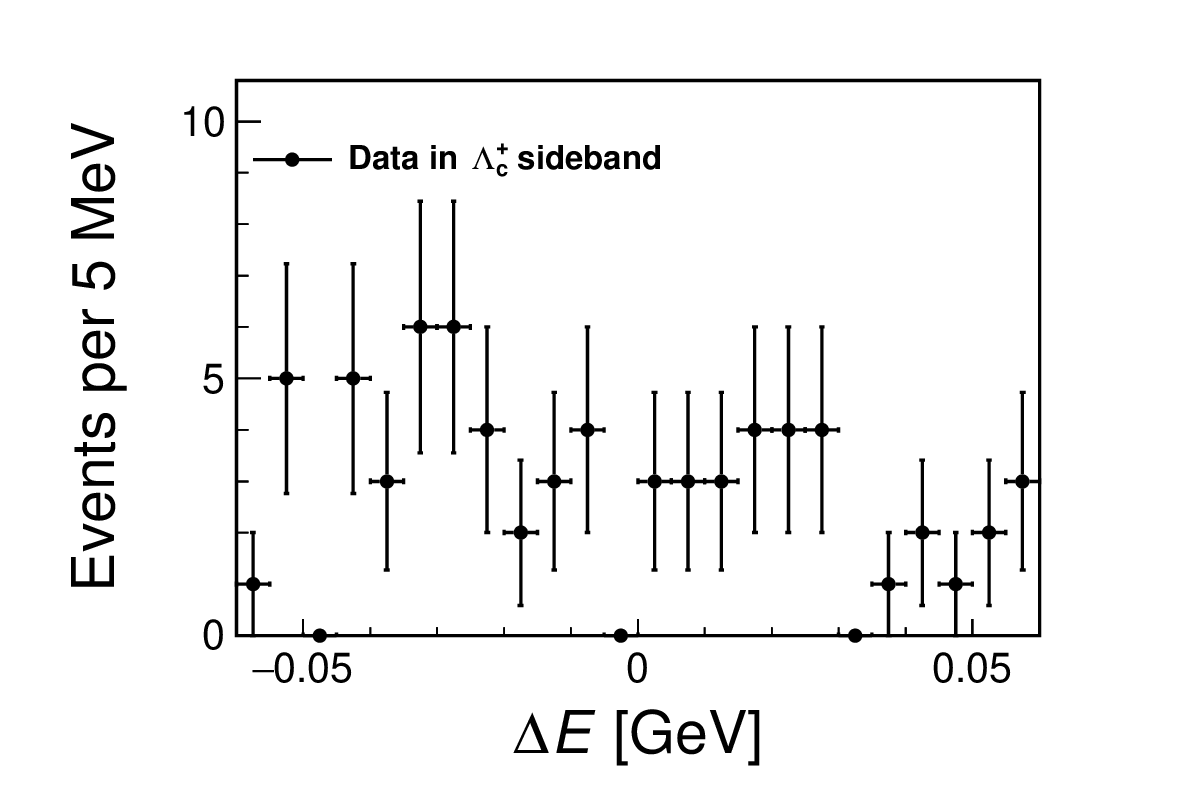}
		\includegraphics[width=7cm]{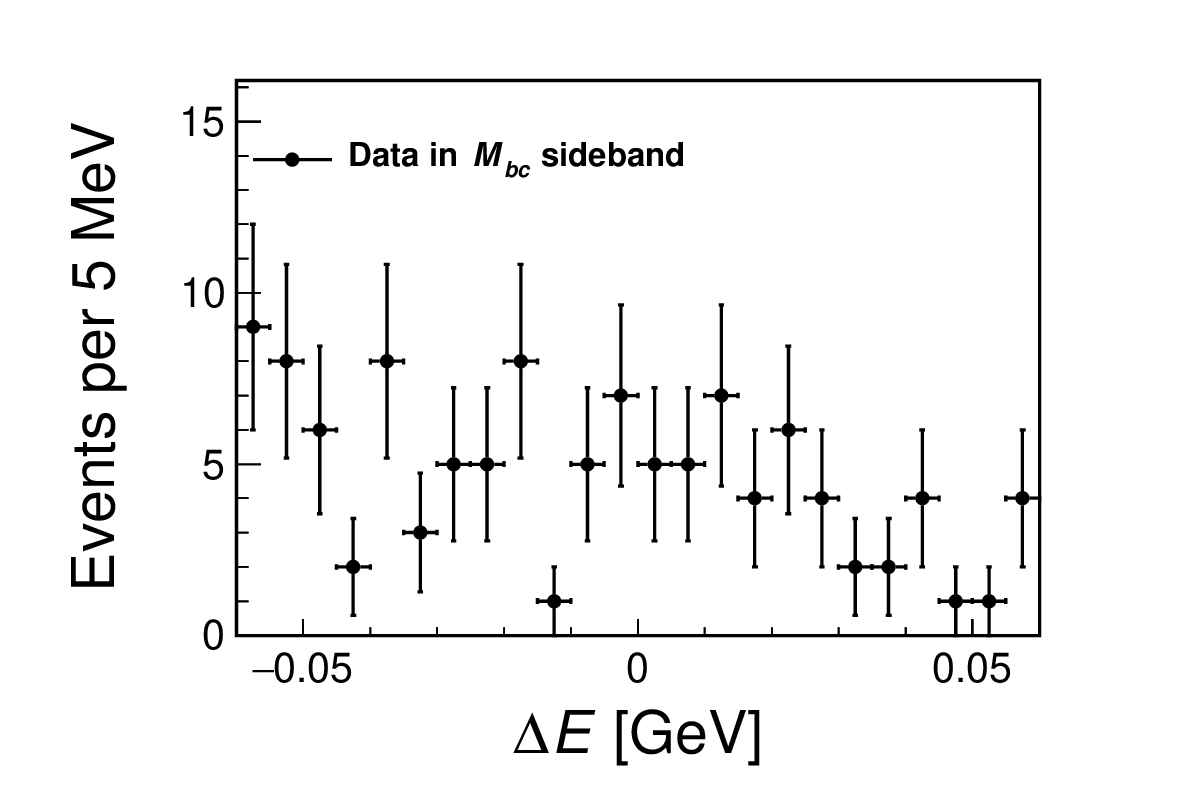}
     	\put(-255,105){\bf (b1)} \put(-53,105){\bf (b2)}
		\caption{Distributions of (a) $M(\Lambda_{c}^{+} \pi^{+})$ and (b) $\Delta E$ from the sideband regions of (1) $M(\Lambda_{c}^{+})$ and  (2) $M_{bc}$
		for the $B^{0} \to \Sigma_{c}(2455)^{0} \overline{\Xi}_{c}^{0}$ decay in the combined Belle and Belle~II data sets.}\label{fig2}
	\end{center}
\end{figure}

\noindent \textit{$M(\Lambda_{c}^{+}\pi^{\pm})$ and $\Delta E$ distributions}: 
Figures~\ref{fig1} and \ref{fig2} show the $M(\Lambda_{c}^{+}\pi^{\pm})$ and $\Delta E$ distributions
derived from the sideband regions of $M(\Lambda_{c}^{+})$ and $M_{\rm bc}$ for  the $B^{+} \to \Sigma_{c}(2455)^{++} \overline{\Xi}_{c}^{-}$ and
$B^{0} \to \Sigma_{c}(2455)^{0} \overline{\Xi}_{c}^{0}$ decays, respectively,  in the combined Belle and Belle~II data sets.
 
\begin{figure}[htbp]
	\begin{center}
		\includegraphics[width=7cm]{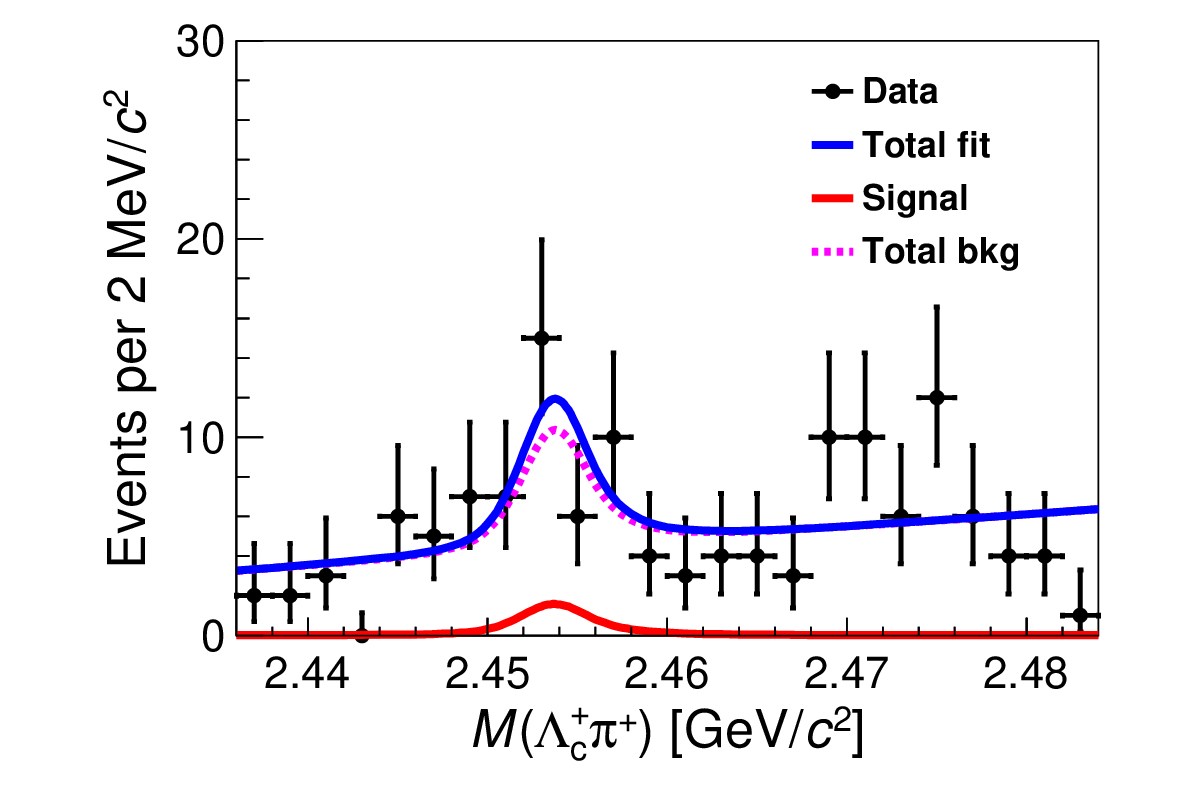}
		\includegraphics[width=7cm]{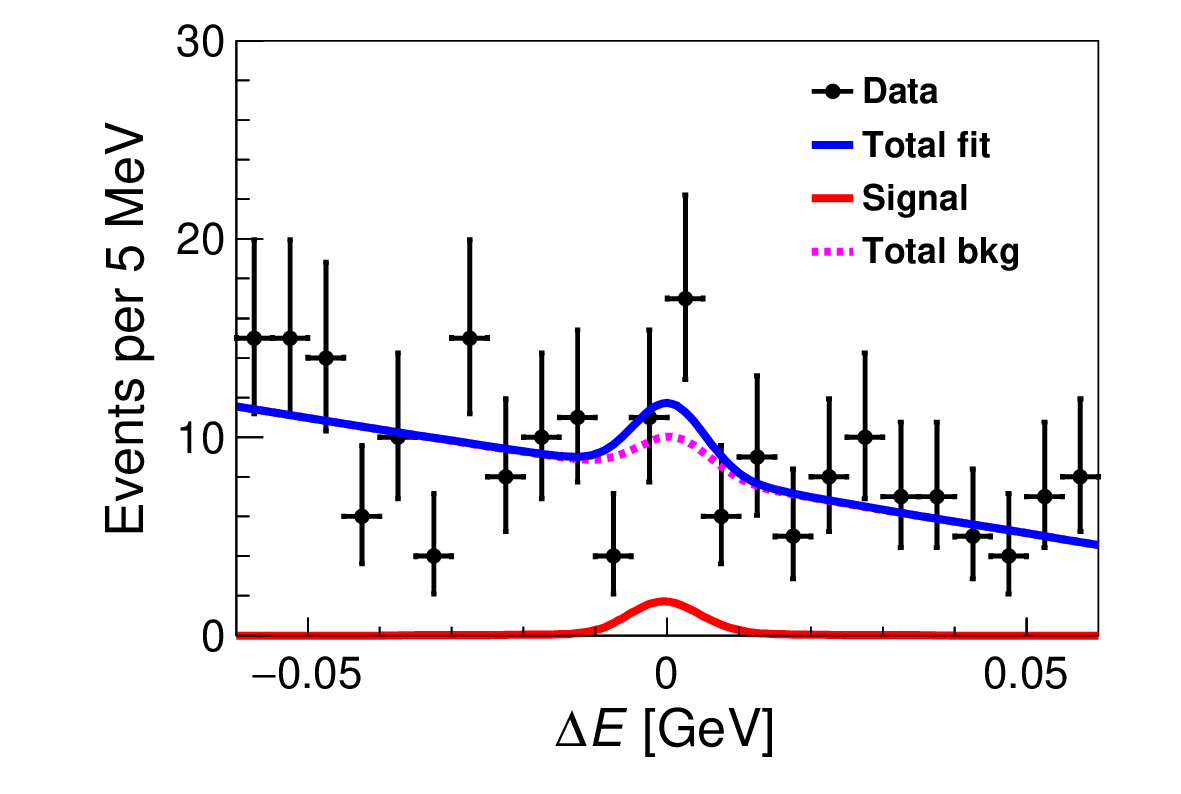}
		\put(-350,115){\bf (a)} \put(-150,115){\bf (b)}
  
		\includegraphics[width=7cm]{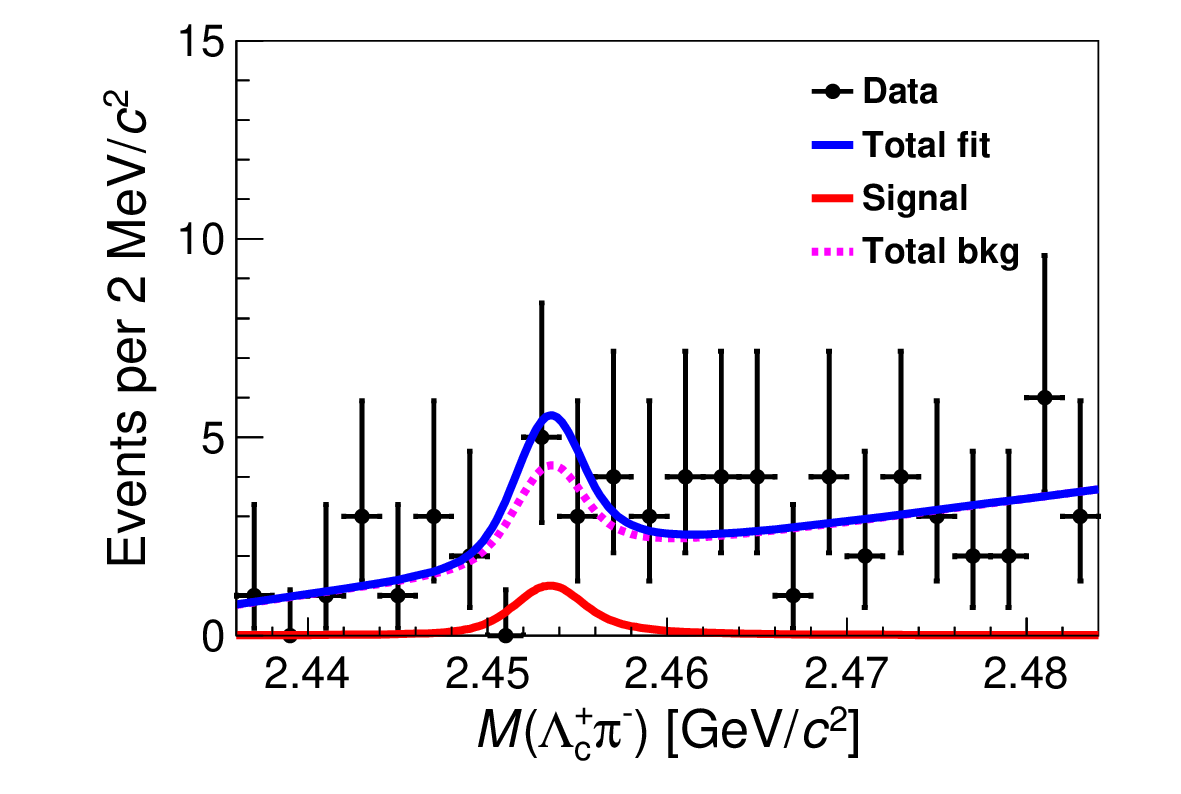}
		\includegraphics[width=7cm]{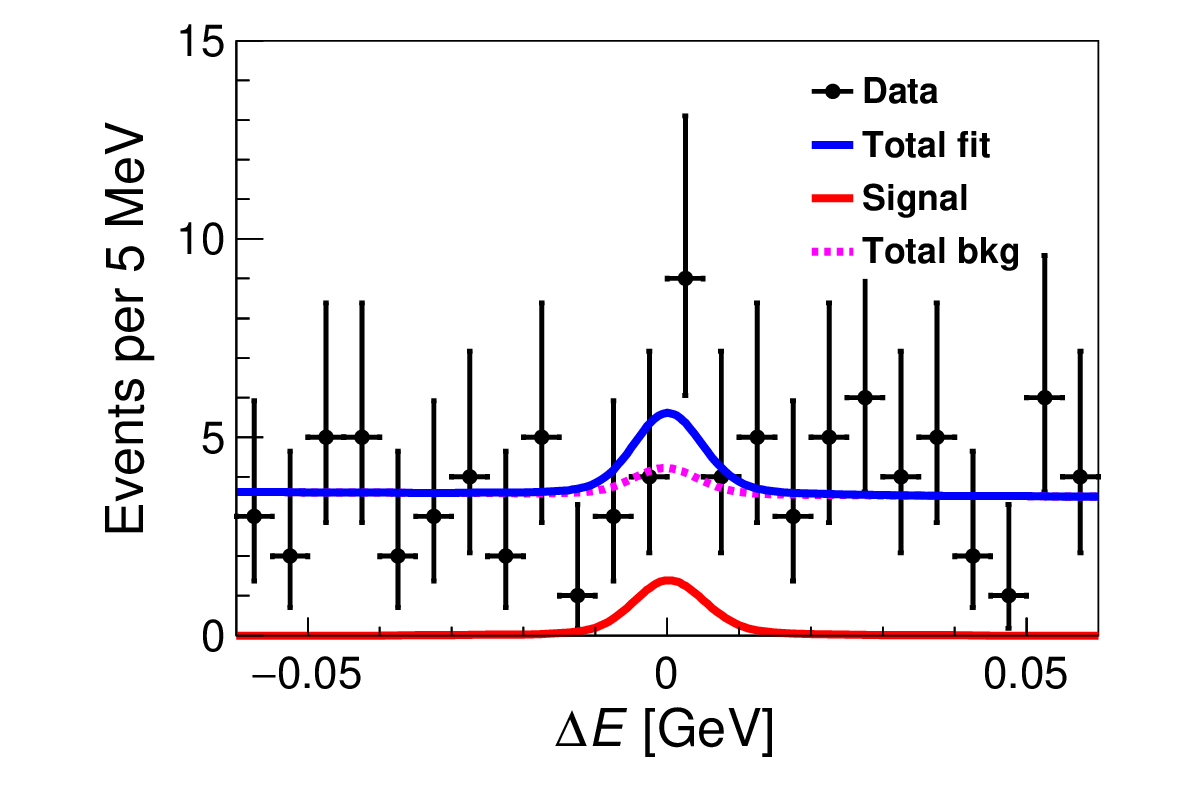}
		\put(-350,115){\bf (c)} \put(-150,115){\bf (d)}
		\caption{Distributions of (a, c) $M(\Lambda_{c}^{+} \pi^{\pm})$ and (b, d) $\Delta E$ for the reconstructed (top) $B^{+} \to \Sigma_{c}(2455)^{++} \overline{\Xi}_{c}^{-}$ 
		and (bottom) $B^{0} \to \Sigma_{c}(2455)^{0} \overline{\Xi}_{c}^{0}$ candidates, using events from the sideband regions of $M(\overline{\Xi}_{c}^{-,0})$ in
		the combined Belle and Belle~II data sets. All components are indicated in the legends.}\label{fig3}
	\end{center}
\end{figure}

\noindent \textit{Fit results to the $M(\Lambda_c^+\pi^\pm)$ and $\Delta E$ distributions from the  sideband regions of $M(\overline{\Xi}_c^{-, 0})$}:
Figure~\ref{fig3} shows the fit results to the $M(\Lambda_c^+ \pi^\pm)$ and $\Delta E$ distributions from the sideband regions 
of $M(\overline{\Xi}_{c}^{-,0})$ in the combined Belle and Belle~II data sets. 

